\documentclass[fleqn,12pt,twoside]{article}
\usepackage{espcrc1}

\usepackage{graphicx}
\usepackage[figuresright]{rotating}

\newcommand{\AmS}{{\protect\the\textfont2
  A\kern-.1667em\lower.5ex\hbox{M}\kern-.125emS}}

\newcommand{\sfrac}[2]{\mbox{\footnotesize $\displaystyle \frac{#1}{#2}$}}
\newcommand{\lsim}{\mathrel{\rlap{\lower4pt\hbox{\hskip0pt$\sim$}} 
\raise1pt\hbox{$<$}}}           
\newcommand{\gsim}{\mathrel{\rlap{\lower4pt\hbox{\hskip0pt$\sim$}} 
\raise1pt\hbox{$>$}}}           

\title{On Nucleon Electromagnetic Form Factors: A Pr\'ecis\thanks{This work was supported by: %
the Austrian Research Foundation \textit{FWF, Erwin-Schr\"o\-din\-ger-Stipendium} no.\ J2233-N08; COSY contract no.\ 41445395; Department of Energy, Office of Nuclear Physics, contract no.\ W-31-109-ENG-38; \textit{Deutsche Forschungsgemeinschaft} contract no.\ GRK683; \textit{Helmholtz-Gemeinschaft} contract no.\ VH-VI-041; National Science Foundation contract no.\ INT-0129236; and the \textit{A.\,v.\ Humboldt-Stiftung} via a \textit{F.\,W.\ Bessel Forschungspreis}.}
}

\author{
A.\ H\"oll,\address[ANL]{%
Physics Division, Argonne National Laboratory, Argonne IL 60439, 
USA} %
R.\ Alkofer,\address[toob]{%
Institut f\"ur Theoretische Physik, Universit\"at T\"ubingen,
Auf der Morgenstelle 14 \\\hspace*{0.3ex} D-72076 T\"ubingen, Germany}
M.\ Kloker,\addressmark[toob]
A.\ Krassnigg,\addressmark[ANL] %
C.\,D.\ Roberts\,\addressmark[ANL]$^{,}$\address{%
Fachbereich Physik, Universit\"at Rostock, D-18051 Rostock, Germany} %
and S.\,V.\ Wright\,\addressmark[ANL] %
}
       
\begin{document}

\maketitle

\begin{abstract}
Electron scattering at large $Q^2$ probes a nucleon's quark core.  This core's contribution to electromagnetic form factors may be calculated using Poincar\'e covariant Faddeev amplitudes combined with a nucleon-photon vertex that automatically fulfills a Ward-Takahashi identity for on-shell nucleons.  The calculated behaviour of $G_E^p(Q^2)/G_M^p(Q^2)$ on $Q^2\in [2,6]\,$GeV$^2$ agrees with that inferred from polarisation transfer data, and exhibits a zero at $Q^2\approx 6.5\,$GeV$^2$.  There is some evidence that $F_2(Q^2)/F_1(Q^2) \propto [\ln Q^2/\Lambda^2]^2/Q^2$ for $Q^2\gsim 6\,$GeV$^2$.
\end{abstract}
\bigskip

The discrepancy between the ratio of electromagnetic proton form factors extracted via Rosenbluth separation and that inferred from polarisation transfer \cite{jones,roygayou,gayou,arrington,qattan} is marked for $Q^2\gsim 2\,$GeV$^2$.  At values of momentum transfer $Q^2 > M^2$, where $M$ is the nucleon's mass, a veracious understanding of such data requires a Poincar\'e covariant description of the nucleon.  That may be obtained via a covariant Faddeev equation \cite{regfe,hugofe}, whose derivation is based on an observation that the same interaction which describes colour-singlet mesons also generates quark-quark (diquark) correlations in the colour-$\bar 3$ (antitriplet) channel \cite{regdq}.  While diquarks do not appear in the strong interaction spectrum \cite{mandarvertex}, the attraction between quarks in this channel supports a picture of baryons in which two quarks are always correlated as a colour-$\bar 3$ diquark pseudoparticle, and binding is effected by the iterated exchange of roles between the bystander and diquark-participant quarks.  The calculation of electromagnetic form factors requires in addition a Ward-Takahashi identity preserving current that is appropriate to a nucleon represented as the solution of the Faddeev equation \cite{oettelpichowsky}.  

Reference~\cite{hoell} proposed that the nucleon is at heart composed of a dressed-quark and nonpointlike diquark.  One element of that study is the dressed-quark propagator.  The form used \cite{mark} both anticipated and expresses features that are now known to be correct \cite{bhagwatmaris}.  It carries no free parameters, because its behaviour was fixed in analyses of meson observables, and is basic to an effective description of light- and heavy-quark mesons \cite{mishasvy}.

\begin{table}[b]
\begin{center}
\caption{Mass-scale parameters (in GeV) for the scalar and axial-vector diquarks, fixed by fitting nucleon and $\Delta$ masses: the fitted mass was offset to allow for ``pion cloud'' contributions \protect\cite{hechtfe}.  $\omega_{J^{P}}= \sfrac{1}{\surd 2}m_{J^{P}}$ is the width-parameter in the $(qq)_{J^P}$-diquark's Bethe-Salpeter amplitude: its inverse is a gauge of the diquark's matter radius.  (Adapted from Ref.\,\protect\cite{hoell}.)\label{ParaFix}}
\begin{tabular*}{1.0\textwidth}{
c@{\extracolsep{0ptplus1fil}}c@{\extracolsep{0ptplus1fil}}|c@{\extracolsep{0ptplus1fil}} c@{\extracolsep{0ptplus1fil}}|c@{\extracolsep{0ptplus1fil}}c@{\extracolsep{0ptplus1fil}}}
\hline
$M_N$ & $M_{\Delta}$~ & $m_{0^{+}}$ & $m_{1^{+}}$~ &
$\omega_{0^{+}} $ & $\omega_{1^{+}}$ \\
1.18 & 1.33~ & 0.79 & 0.89~ & 0.56=1/(0.35\,{\rm fm}) & 0.63=1/(0.31\,{\rm fm}) \\
\hline
\end{tabular*}
\end{center}
\end{table}

The nucleon bound state is subsequently realised via a Poincar\'e covariant Faddeev equation, which incorporates scalar and axial-vector diquark correlations.  In this there are two parameters: the mass-scales associated with the correlations.  They were fixed by fitting to specified nucleon and $\Delta$ masses: the values are listed in Table~\ref{ParaFix}.  The study thus arrived at a representation of the nucleon that possesses no free parameters with which to influence the nucleons' form factors.

\begin{figure}[t]
\vspace*{-4ex}

\centerline{
\includegraphics[width=0.63\textwidth,angle=270]{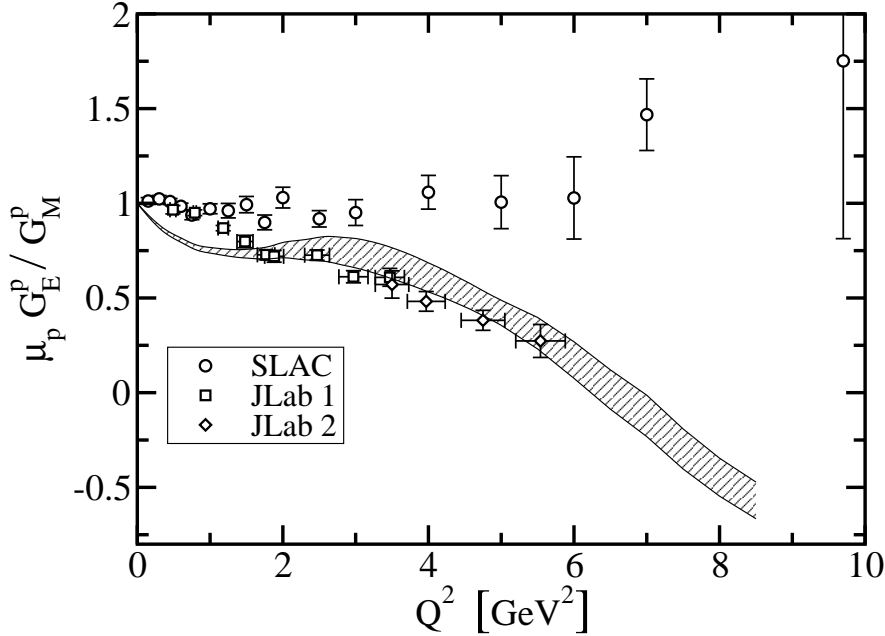}}
\vspace*{-4ex}

\caption{\label{plotGEpGMp} Form factor ratio: $\mu_p\, G_E^p(Q^2)/G_M^p(Q^2)$.  $G_E^p(Q^2)$ was calculated using the point-particle values: $\mu_{1^+}=2$ \& $\chi_{1^+}=1$, and $\kappa_{\cal T} = 2$.  Variations in the axial-vector diquark parameters used to evaluate $G_E^p(Q^2)$ have little effect on the results.  The width of the band reflects the variation in $G_M^p(Q^2)$ with axial-vector diquark parameters and, in both cases, the upper border is obtained with $\mu_{1^+}=3$, $\chi_{1^+}=1$ and $\kappa_{\cal T}= 2$, while the lower has $\mu_{1^+}= 1$.  The data are: \textit{squares} - Ref.\,\cite{jones}; \textit{diamonds} - Ref.\,\cite{gayou}; and \textit{circles} - Ref.\,\cite{walker}.
}
\end{figure}

At this point only a specification of the nucleons' electromagnetic interaction remained.  Its formulation was primarily guided  by a requirement that the nucleon-photon vertex satisfy a Ward-Takahashi identity.  The result depends on three parameters tied to properties of the axial-vector diquark correlation: $\mu_{1^+}$ \& $\chi_{1^+}$, respectively, the axial-vector diquarks' magnetic dipole and electric quadrupole moments; and $\kappa_{\cal T}$, the strength of electromagnetic axial-vector $\leftrightarrow$ scalar diquark transitions.  

In Fig.\,\ref{plotGEpGMp} we plot a ratio of the proton's Sachs electric and magnetic form factors; viz., $\mu_p\, G_E^p(Q^2)/G_M^p(Q^2)$.   The behaviour of the data at small $Q^2$ is readily understood.  In the neighbourhood of $Q^2=0$, 
\begin{equation}
\mu_p\,\frac{ G_E^p(Q^2)}{G_M^p(Q^2)} = 1 - \frac{Q^2}{6} \,\left[ (r_p)^2 - (r_p^\mu)^2 \right]\,,
\end{equation}
and because $r_p\approx r_p^\mu$; viz., the proton's electric and magnetic radii are approximately equal, the ratio varies by less than 10\% on $0<Q^2< 0.6\,$GeV$^2$, if the form factors are approximately dipole.  This is evidently true of the experimental data.  The calculated curve was obtained ignoring the contribution from pion loops.  Without such chiral corrections, $r_p> r_p^\mu$ and hence the calculated ratio falls immediately with increasing $Q^2$.  Incorporating pion loops one readily finds $r_p\approx r_p^\mu$ \cite{hoell}.  It is thus apparent that the small $Q^2$ behaviour of this ratio is materially affected by the proton's pion cloud.

Pseudoscalar mesons are not pointlike and therefore pion cloud contributions to form factors diminish in magnitude with increasing $Q^2$.  The evolution of $\mu_p\, G_E^p(Q^2)/G_M^p(Q^2)$ on $Q^2\gsim 2\,$GeV$^2$ is primarily determined by the quark core of the proton.  This is evident in Fig.\,\ref{plotGEpGMp}, which illustrates that for $Q^2 \gsim 2 \,$GeV$^2$, $\mu_p\, G_E^p(Q^2)/G_M^p(Q^2)$ is sensitive to the parameters defining the axial-vector-diquark--photon vertex.  The ratio passes through zero at $Q^2 \approx 6.5\,$GeV$^2$; namely, at the point for which $G_E^p(Q^2)=0$.  In this model the existence of the zero is robust but its location depends on the model's parameters.  In comparison with Ref.\,\cite{hoell}, this figure was obtained with increased numerical accuracy in the calculation of the Faddeev amplitudes and form factors, and expands the domain of the calculation.  The behaviour of $\mu_p\, G_E^p(Q^2)/G_M^p(Q^2)$ owes itself primarily to spin-isospin correlations in the nucleon's Faddeev amplitude.

\begin{figure}[t]
\vspace*{-4ex}

\centerline{\hspace*{2em}%
\includegraphics[width=0.63\textwidth,angle=270]{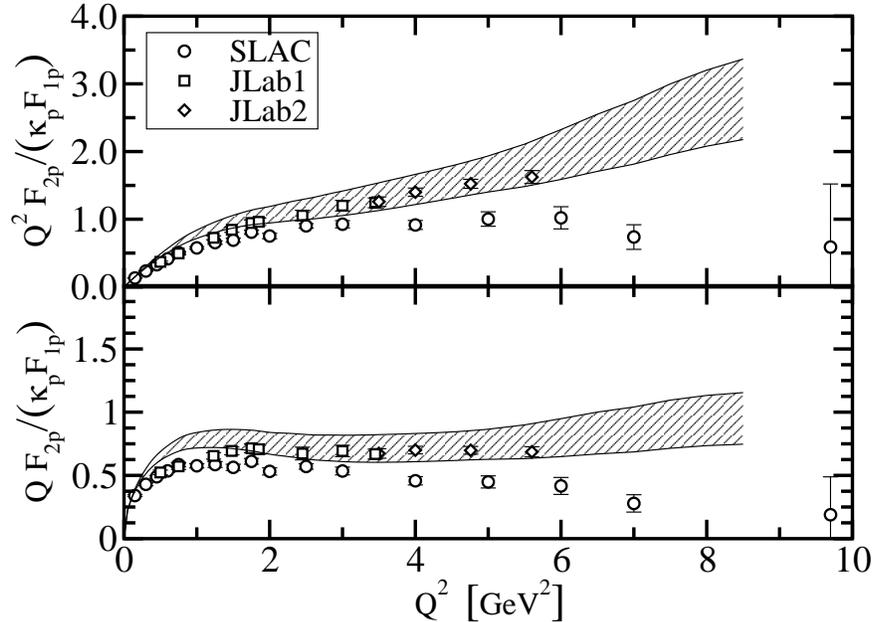}}
\vspace*{-4ex}

\caption{\label{plotF2F1} Proton Pauli$/$Dirac form factor ratios.  The data are as described in Fig.\,\protect\ref{plotGEpGMp}, as is the band except that here the upper border is obtained with $\mu_{1^+}=1$, $\chi_{1^+}=1$ and $\kappa_{\cal T}= 2$, and the lower with $\mu_{1^+}=3$.}
\end{figure}

In Fig.\,\ref{plotF2F1} we depict a weighted ratio of the proton's Dirac and Pauli form factors.  The numerical results are consistent with 
\begin{equation}
 \sqrt{Q^2}\,\frac{ F_2(Q^2)}{F_1(Q^2)}\; 
 \approx {\rm constant},\; 2 \lsim  Q^2 ({\rm GeV}^2) \lsim 6\,,
\end{equation}
as are the polarisation transfer data.  However, the present calculation hints that this scaling relation fails for $Q^2 \gsim 6\,$GeV$^2$.

\begin{figure}[t]
\vspace*{-4ex}

\centerline{\hspace*{2em}%
\includegraphics[width=0.63\textwidth,angle=270]{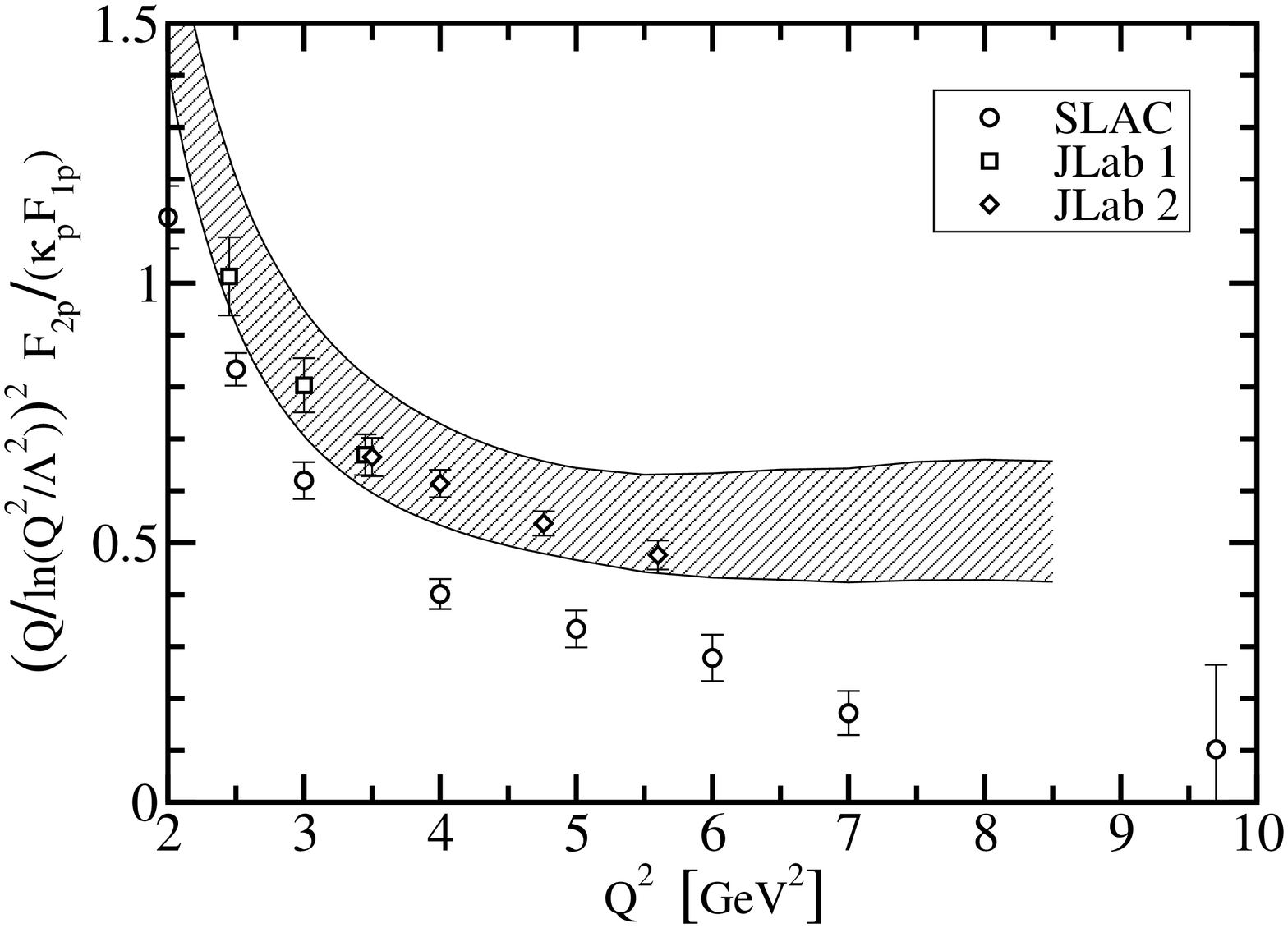}}
\vspace*{-4ex}

\caption{\label{plotF2F1log}  Weighted proton Pauli$/$Dirac form factor ratio, calculated with $\Lambda= 0.94\,$GeV.  The band is as described in Fig.\,\protect\ref{plotF2F1}, as are the data.}
\end{figure}

In Fig.\,\ref{plotF2F1log} we plot another weighted ratio of Pauli and Dirac form factors.  A perturbative QCD analysis \cite{belitsky} that considers effects arising from both the proton's leading- and subleading-twist light-cone wave functions, the latter of which represents quarks with one unit of orbital angular momentum, suggests
\begin{equation}
\label{scaling}
\frac{Q^2}{[\ln Q^2/\Lambda^2]^2} \, \frac{F_2(Q^2)}{F_1(Q^2)} =\,{\rm constant,}\;\; Q^2\gg \Lambda^2\,,
\end{equation}  
where $\Lambda$ is a mass-scale that corresponds to an upper-bound on the domain of soft momenta.  An argument may be made that a judicious estimate of the least-upper-bound on this domain is $\Lambda = M$ \cite{hoell}.  The figure hints that Eq.\,(\ref{scaling}) may be valid for $Q^2 \gsim 6\,$GeV$^2$.

It is noteworthy that orbital angular momentum is not a Poincar\'e invariant.  However, if absent in a particular frame, it will almost inevitably appear in another frame related via a Poincar\'e transformation.  Nonzero quark orbital angular momentum is a necessary outcome of a Poincar\'e covariant description.  This is why a nucleon's covariant Faddeev amplitude is a matrix-valued function with a rich structure that, in the nucleons' rest frame, corresponds to a relativistic wave function with $s$-wave, $p$-wave and even $d$-wave components \cite{oettel}.  The result in Fig.\,\ref{plotF2F1log} is not significantly influenced by details of the diquarks' electromagnetic properties.  Instead, the behaviour is primarily governed by correlations expressed in the proton's Faddeev amplitude and, in particular, by the amount of intrinsic quark orbital angular momentum \cite{blochff}.  



\end{document}